\documentclass[%
 reprint,
 amsmath,amssymb,
 aps,
]{revtex4-2}

\usepackage{float}
\usepackage{graphicx}
\usepackage{dcolumn}
\usepackage{bm}

\usepackage{hyperref}

\begin{document}

\author{Uba K. Ubamanyu}%
\affiliation{%
    Institute of Mechanical Engineering,\\
    \'{E}cole Polytechnique F\'{e}d\'{e}rale de Lausanne (EPFL),\\
    1015 Lausanne, Switzerland\\
}%

\author{Zheren Baizhikova}
\affiliation{Department of Civil and Environmental Engineering,\\ 
    University of Houston,\\
    Houston, TX 77204, USA\\
}

\author{Jia-Liang Le}
\affiliation{Department of Civil, Environmental, and Geo Engineering,\\ 
    University of Minnesota,\\
    Minneapolis, MN 55455, USA\\
}

\author{Roberto Ballarini}
\affiliation{Department of Civil and Environmental Engineering,\\ 
    University of Houston,\\
    Houston, TX 77204, USA\\
}

\author{Pedro M. Reis}%
 \email{pedro.reis@epfl.ch}
\affiliation{%
    Institute of Mechanical Engineering,\\
    \'{E}cole Polytechnique F\'{e}d\'{e}rale de Lausanne (EPFL),\\
    1015 Lausanne, Switzerland\\
    }%


\title{A numerical study on the buckling of near-perfect spherical shells}

\keywords{Spherical shells; Buckling; Imperfection Sensitivity; Knockdown factor; Finite-Element Simulations}

\begin{abstract}
We present the results from a numerical investigation using the finite element method to study the buckling strength of near-perfect spherical shells containing a single, localized, Gaussian-dimple defect whose profile is systematically varied toward the limit of vanishing amplitude. In this limit, our simulations reveal distinct buckling behaviors for hemispheres, full spheres, and partial spherical caps. Hemispherical shells exhibit boundary-dominated buckling modes, resulting in a knockdown factor of 0.8. By contrast, full spherical shells display localized buckling at their pole with knockdown factors near unity. Furthermore, for partial spherical shells, we observed a transition from boundary modes to these localized buckling modes as a function of the cap angle. We characterize these behaviors by systematically examining the effects of the discretization level, solver parameters, and radius-to-thickness ratio on knockdown factors. Specifically, we identify the conditions under which knockdown factors converge across shell configurations. Our findings highlight the critical importance of carefully controlled numerical parameters in shell-buckling simulations in the near-perfect limit, demonstrating how precise choices in discretization and solver parameters are essential for accurately predicting the distinct buckling modes across different shell geometries.
\end{abstract}

\maketitle 


\section{Introduction}
\label{sec_Intro}

Studying the buckling of spherical shells has a long history due to their structural efficiency. However, predicting their buckling strength remains challenging because of sensitivity to imperfections. Experiments on realistic shells, which are inevitably imperfect, have evidenced buckling loads as low as 20\% \textit{vis-\`a-vis} theoretical predictions, calling for the use of knockdown factors~\cite{thurston_effect_1966, carlson_experimental_1967, weingarten_buckling_1968, hilburger_buckling_2020}. The knockdown factor, $\kappa$, is defined as the ratio between experimentally (or numerically) measured critical buckling loads and the corresponding theoretical prediction. 
The classical theoretical buckling pressure of a perfect spherical shell under external pressure, derived by Zoelly in 1915 using linear buckling analysis~\cite{zoelly_ueber_1915}, is commonly used to normalize the buckling pressure:
\begin{equation}
\label{Eq_Zoelly}
p_c = \frac{2E}{\sqrt{3(1-\nu^2)}}\left(\frac{t}{R}\right)^{2}
\end{equation}
where, $E$ is the Young's modulus and $\nu$ is the Poisson's ratio of the material, $R$ is the radius of the sphere, and $t$ is the thickness of the shell.  The prevalence of knockdown-factor values below unity ($\kappa{<}1$) is widely attributed to geometric imperfections, either introduced during manufacturing or resulting from in-service conditions. This phenomenon is well-documented, with extensive experimental, computational, and theoretical literature spanning nearly 85 years. For a more comprehensive overview of the relevant literature, we direct the reader to the seminal studies in the following Refs.~\cite{von_karman_buckling_1939, koiter_over_1945, budiansky_buckling_1959, huang_unsymmetrical_1964, hutchinson_imperfection_1967, kaplan_buckling_1974, bushnell_computerized_1985, elishakoff_resolution_2014, hutchinson_buckling_2016}.

Early numerical studies~\cite{kaplan_nonlinear_1954, budiansky_buckling_1959, hutchinson_postbuckling_1970, bushnell_computerized_1981} provided valuable insights into the imperfection sensitivity of both full and partial spherical shells, but computational limitations often constrained their scope. With recent advancements in computation power and precision, it is now feasible to undertake more extensive parameter studies, capturing a broader range of geometric imperfections, shell configurations, and loading conditions.
From past studies, it is well-established that knockdown factors decrease towards a plateau as the defect amplitude increases, suggesting, in reverse, that knockdown factors should approach unity as shells near the perfect geometry~\cite{koiter_over_1945, hutchinson_imperfection_1967, lee_geometric_2016, jimenez_technical_2017}. However, the scarcity of experimental and numerical data showing knockdown factors above $\kappa{>}0.9$ suggests that additional effects or numerical artifacts beyond just geometric imperfections may be at play for near-perfect shells. 

The effect of boundary conditions on shallow spherical shells has been extensively studied, though early analytical findings often conflicted with experimental observations. Early analyses were limited to symmetric buckling modes, failing to capture the asymmetric buckling behaviors observed in experiments. Kaplan consolidated a substantial body of experimental, analytical, and numerical results in the chapter "\textit{Buckling of Spherical Shells}" in Ref.\cite{kaplan_buckling_1974}, highlighting the dependence of buckling behavior on the shallowness parameter $\lambda_S$ (Eq.\,\eqref{Eq_lambdaS} defined in \S\ref{sec_probdef}). For shallow shells with lower $\lambda_S$ values, symmetric buckling dominates, making symmetric theories applicable in this regime. However, as $\lambda_S$ increases, the buckling transitions to asymmetric modes called for asymmetric buckling theories.
Using a two-term Galerkin method, Parmerter and Fung \cite{parmerter_influence_1962} demonstrated that asymmetric buckling is likely to occur when $\lambda_S > 5.5$, a threshold later confirmed by Huang \cite{huang_unsymmetrical_1964} using finite difference methods. Further analytical studies \cite{parmerter_buckling_1964, thurston_asymmetrical_1964} supported this limit as the transition between symmetric and asymmetric buckling modes. It is important to highlight that this transition to asymmetrical buckling modes necessitates a full 3D modeling approach, whereas the axisymmetric models fail to capture such a phenomenon.

Understanding the influence of imperfections on knockdown factors is essential for accurately predicting shell buckling, especially as manufacturing capabilities now enable the production of high-precision shells. While prior studies mentioned above have primarily focused separately on the role of geometric defects and boundary conditions, the precise influence of boundary conditions and geometric defects causing different buckling modes in near-perfect shells remains poorly understood. 

Here, we conduct a series of numerical simulations using the finite element method (FEM) to investigate the validity and limitations of imperfection sensitivity of the buckling onset due to a geometric defect in the limit of near-perfect spherical shells. We consider a range of defect amplitudes and shell configurations, gradually reducing defect amplitude to near zero, so as to closely approach near-perfect shell conditions. With the recent advances in computation power, one could be inclined to normalize the observed buckling pressure by the buckling pressure obtained using the FEM simulations of perfect geometry instead of theoretical predictions; such Eq.~(\ref{Eq_Zoelly}) derived by Zoelly for spherical shells. Doing so may be particularly tempting for non-spherical shells for which there may be no close-formed analytical solutions.
A key goal of this study, while focusing on spherical shells, is to assess the impact of discretization, solver parameters, and geometry in predicting the buckling response of near-perfect shells, including both full and partial geometries. We highlight the potential risks of simulating idealized geometries without imperfections and demonstrate that FEM simulations for these geometries yield results that are highly sensitive to numerical parameters. Thus, our findings provide a recommendation against using the simulated perfectly spherical shell case to normalize the buckling pressures. Normalization using classical prediction provides a unifying framework for comparing results across studies and facilitates a consistent evaluation of imperfection effects in spherical shells. Furthermore, by examining both deep and shallow partial spherical caps with varying cap half-angles, we capture the distinct boundary-dominated buckling modes that emerge in partial shells.

Our manuscript is structured as follows: \S\ref{sec_probdef} 
defines the problem at hand, and \S\ref{sec_FEM} details the numerical methodology followed to tackle it. In \S\ref{sec_sensitivity}, we examine the effects of discretization and solver parameters on the FEM results, followed, in \S\ref{sec_boundary_modes}, by a discussion of the distinct buckling modes observed. \S\ref{sec_effect_of_eta} explores the influence of the radius-to-thickness ratio on imperfection sensitivity, and \S\ref{sec_shallow} presents the buckling behavior of partial spherical shells. Finally, \S\ref{sec_conclusion} concludes with a discussion of the key findings and highlights directions for future work.

\section{Problem definition}
\label{sec_probdef}

\begin{figure}[hbt!]
\centering\includegraphics{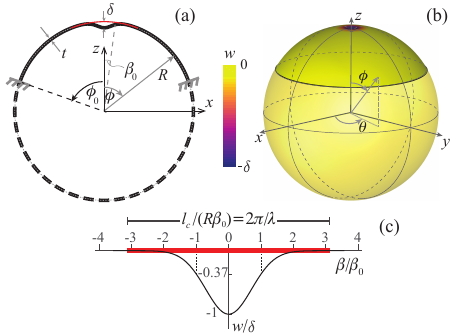}
\caption{Geometry of a typical spherical shell with a Gaussian defect at the pole, clamped at its free boundary. 
(a) Meridian cross-section of the shell, defining all the relevant geometric parameters.
(b) 3D visualization of a partial shell in the spherical coordinate system; the color map represents the radial deviation, $w$, from a perfect sphere. 
(c) Parameterized Gaussian defect profile $w(\beta)$ according to Eq.~(\ref{Eq_Gaussian}), for $\lambda_I=1$. The red line represents angular width, $l_c/(R\beta_0)$, associated with the theoretical buckling wavelength $l_c$ (see text) of the axisymmetric mode for this particular shell.
\label{fig1_ProblemDef}}
\end{figure}

We examine three distinct types of thin spherical geometries of shells subjected to external pressure: (i) a hemispherical shell clamped along its equator, (ii) a full spherical shell, and (iii) partial spherical shells, from shallow to deep, in a range of cap half-angles denoted by $\phi_0$; see Fig.~\ref{fig1_ProblemDef}(a). Without loss of generality, we set the radius of the undeformed middle surface of the shell to $R {=} 20\,$mm and the shell thickness to $t {=} 0.2\,$mm, resulting in a radius-to-thickness ratio of $R/t {=} 100$. 
Additionally, we will conduct simulations across a range of radius-to-thickness ratios,
$50 {\leq} R/t{\leq} 1500$, to attest the robustness and generality of the results across various geometries. For the partial spherical shells, $\phi_0$ is varied from 15$^\circ$ (shallow shell) to 180$^\circ$ (full shell), each clamped along the free boundaries. Note that a full spherical shell does not have a boundary to be clamped, given its fully closed geometry, and the model is in static self-equilibrium due to the uniformity of the external pressure applied throughout its surface.

The shallowness of the shell is characterized~\cite{ kaplan_nonlinear_1954, budiansky_buckling_1959, huang_unsymmetrical_1964} by the geometric parameter
\begin{equation}
    \label{Eq_lambdaS}
    \lambda_S = [12(1-\nu^2)]^{1/4}\sqrt{R/t}\phi_0.
\end{equation}
where, $\nu$ is the Poisson's ratio of the material.
In the literature, this parameter has been defined with slight variations using the shallow shell's base radius or height. For sufficiently shallow shells, all these variations are equivalent. However, since we are interested in studying both deep and shallow partial spherical shells, we define $\lambda_S$ in terms of the cap half-angle, $\phi_0$.
For convenience, this parameter can be interpreted in terms of the ratio between the arclength of the shell and the theoretical buckling wavelength of an axisymmetric mode of the spherical shell, $l_c$~\cite{hutchinson_imperfection_1967}, where $l_c = 2\pi[12(1-\nu^2]^{-1/4}\sqrt{Rt}$; thus, $\lambda_S = \pi 2R\phi_0/l_c$. For the chosen geometry with $R/t = 100$ and $\nu = 0.3$, the buckling wavelength is $l_c\approx 6.9\,$mm. When $\lambda_S = \pi$, one full wavelength of the axisymmetric buckling mode fits along the arclength of the cap, a critical threshold below which snap-through behavior is absent~\cite{ki_combined_2024}. Additionally, Hutchinson~\cite{hutchinson_imperfection_1967} has suggested that imperfection effects and the post-buckling behavior described by Koiter’s general theory apply reliably only for shallow spherical caps with $\lambda_S > 3\pi$.

All the shell geometries that we will consider incorporate a precisely defined geometric imperfection: a Gaussian-shaped dimple located at the pole. 
Meridional cross-section and three-dimensional (3D) representations of the undeformed shell geometry are illustrated in Fig.~\ref{fig1_ProblemDef}(a) and (b), respectively. A spherical coordinate system $(r,\, \theta,\, \phi)$ is used, where $r$ is the radial distance from the origin/center of the sphere, $\theta$ is the circumferential angle, and $\phi$ is the meridian angle, ranging from $\phi=0$ at the pole to $\phi=\phi_0$ at the clamped boundary. 
In Cartesian coordinates $(\textbf{e$_x$},\,\textbf{e$_y$},\,\textbf{e$_z$})$, the position vector of the mid-surface of the undeformed shell with a Gaussian defect is,
$\mathbf{r} = (R + w)\sin{\phi}\cos{\theta}\,{\textbf{e$_x$}} + (R + w)\sin{\phi}\sin{\theta}\,{\textbf{e$_y$}} + (R + w)\cos{\phi}\,{\textbf{e$_z$}}$, where $w$ is the radial deviation from the mid-surface of the perfect sphere. 
This radial deviation associated with the Gaussian defect is:
\begin{equation}
\label{Eq_Gaussian}
w(\beta) = -\delta\ \mathrm{e}^{-\left(\frac{\beta}{\beta_0}\right)^{2}}
\end{equation}
where $\beta$ is the zenith angle from the center of the defect, $\delta$ is the amplitude of the defect, and $\beta_0$ is the characteristic angular half-width parameter.

In line with previous studies~\cite{koga_axisymmetric_1969, lee_fabrication_2016, hutchinson_buckling_2016, jimenez_technical_2017}, we normalize the defect amplitude as $\bar{\delta}=\delta/t$. Additionally, the defect width parameter $\lambda_I = [12(1-\nu^2)]^{1/4}\sqrt{R/t}\beta_0$ is introduced to define the characteristic width of the dimple. Note that this parameter closely resembles the shallowness parameter $\lambda_S$ defined in Eq.~(\ref{Eq_lambdaS}),  though $\lambda_I$ uses the defect’s characteristic angular half-width $\beta_0$, in contrast with $\phi_0$ for $\lambda_S$. For all the cases we will investigate, we fix $\lambda_I=1$, while the normalized defect amplitude varies in the range $0.001\leq\Bar{\delta}\leq5$, allowing the analysis to approach the near-perfect geometry, in the limit of $\Bar{\delta}\rightarrow 0$. The near-perfect limit is defined, somewhat \textit{ad hoc}, as $\Bar{\delta}\leq0.1$.
Figure \ref{fig1_ProblemDef}(c) shows the parameterized Gaussian dimple shape for $\lambda_I = 1$, with the horizontal (red) line indicating the angular extent of a single theoretical buckling wavelength $l_c$.

The spherical shells are loaded with a uniform external pressure applied on the outer surface until the onset of buckling induced by the ensuing compressive stresses. The observed critical buckling pressure, $p_{\text{crit}}$, is identified as the first peak pressure observed at the buckling onset during the loading. 
The knockdown factor is defined as $\kappa = p_{\text{crit}}/p_c$, where $p_c$ is the theoretical buckling pressure of a perfect shell already given in Eq.~\ref{Eq_Zoelly}. 

Our primary objective is to quantify the knockdown factor, $\kappa$, for near-perfect shells in the limit of imperfections with vanishing amplitude, $\Bar{\delta}\rightarrow 0$.
We seek to answer the following questions: How do simulation parameters such as discretization and solver parameters affect the fidelity of predictions for the buckling strength of near-perfect shells? Is there a difference in how the imperfection size and boundary conditions affect the actual buckling behavior of near-perfect partial \textit{vs.} full spherical shells? And finally, how do the radius-to-thickness ratio and shell shallowness influence imperfection sensitivity?

\section{Methodology: Finite element simulations} \label{sec_FEM}

We have conducted FE simulations by using the commercial software package Abaqus/Standard (v2023). In our numerical model, the shell was represented by its 3D mid-surface, which was discretized using four-node, reduced-integration shell elements (S4R). With these choices, the buckling behavior of the pressurized shell can be accurately captured while ensuring computational efficiency, thereby enabling a systematic exploration of the parameter space. Our modeling approach follows the methodology established in previous studies~\cite{abbasi_comparing_2023, derveni_defect-defect_2023, derveni_probabilistic_2023}, where FEM simulations were validated thoroughly against experimental results, ensuring confidence in the accuracy and reliability of the numerical results. Note that even earlier works from our group~\cite{lee_geometric_2016, jimenez_technical_2017,marthelot_buckling_2017, pezzulla_weak_2019, yan_buckling_2020, pezzulla_weak_2019} were restricted to axisymmetric conditions, which we relax here to be able to consider non-axisymmetric modes.

Starting from the perfect shell geometry, the initial geometric imperfection was introduced as a Gaussian dimple at the pole with the profile in Eq.~\eqref{Eq_Gaussian}, specifying the normalized amplitude, $\bar{\delta}$ and angular width, $\lambda_I$, of the defect. This imperfection was applied directly to the mesh by radially adjusting the coordinates of the middle surface of the otherwise perfect geometry.

A linear elastic material model with $E = 210\,$GPa, and $\nu = 0.3$ was used for all the simulations. These material parameters were chosen to match a recent study~\cite{baizhikova_uncovering_2024}. However, as confirmed below in \S\ref{sec_sensitivity}, where we will establish a comparison with earlier results from Ref.~\cite{lee_geometric_2016} (different material parameters), the imperfection sensitivity of a thin spherical shell is independent of the material properties, given the elastic buckling conditions. Thus, the results of this study can be used for other material models without loss of generality. An external live pressure, equivalent to the classical theoretical buckling pressure in Eq.~\eqref{Eq_Zoelly}, was applied uniformly on the outer surface of the shell. The analysis was performed using the Static/Riks method, an arclength-based procedure in Abaqus/Standard that is well-suited for simulating unstable buckling paths as it simultaneously solves for loads and deformations by progressing along the arclength of the load-deformation curve~\cite{riks_incremental_1979}.

A detailed mesh sensitivity analysis was performed to determine an optimal mesh discretization that ensures solution convergence while minimizing computational costs; these results are an integral part of our investigation and are presented below in \S\ref{sec_sensitivity}. We define the discretization level as $m = \pi R/(2d_m)$, representing the number of finite elements along the quarter meridian from the pole to the equator, where $d_m$ is the approximate element size. As detailed in \S\ref{sec_sensitivity} and \S\ref{sec_boundary_modes}, we explored the discretization level in the range $m\in[30,\,200]$. For our specific spherical geometry with $R = 20\,$mm, we have identified a satisfactory nominal mesh size of $d_m\approx0.2\,$mm, corresponding to $m = 150$ elements along the quarter meridian of the shell to simulate converged results in the near-perfect limit. Beyond \S\ref{sec_boundary_modes}, for consistency, only this nominal value will be used for the discretization level.

In addition to systematically exploring $m$, we also conducted a sensitivity analysis on the arclength increment parameters within the Static/Riks method: initial $\Delta s_{ini}$, minimum $\Delta s_{min}$, and maximum $\Delta s_{max}$ arclength increments. The goal was to identify satisfactory values for these parameters that ensure numerical stability and convergence. We explored the maximum admissible increment in the range $\Delta s_{max}\in[0.01,\,0.1]$, while the initial and minimum arclength increment were set to $\Delta s_{ini} = 0.01$ and $\Delta s_{min} = 10^{-10}$, respectively. For our specific geometry, we have identified a maximum admissible arclength increment $\Delta s_{max} = 0.02$, which ensures stability during solution progression. It is important to note that the arclength increments depend on the nominal loading applied. Therefore, for future studies, the classical buckling pressure in Eq.~\eqref{Eq_Zoelly} should be used as the nominal load to ensure the arclength solver accurately computes the load proportionality factor at each increment. 

The results of both the mesh sensitivity and solver parameter analyses, including convergence trends and optimal parameter choices, will be presented in \S \ref{sec_sensitivity}, specifically in Fig.~\ref{fig2_senstivity}. 

In \S\ref{sec_shallow}, beyond spherical and hemispherical shells, we extended our analysis to other partial spherical geometries with a range of cap half-angles spanning from shallow ($0<\phi_0<\pi/2$) to deep ($\pi/2<\phi_0<\pi$) shells. These partial shell models were constructed following the same FEM procedure used for the hemispherical shells, with each cap clamped at its free boundary and subjected to uniform external pressure. For this part of the study, to maintain consistency in discretization and solver performance, the mesh density was kept at a nominal size of $d_m=0.2\,$mm, with an admissible maximum arclength increment of $\Delta s_{max}=0.02$ across all simulations. 

\section{Effect of discretization and solver parameters} \label{sec_sensitivity}

\begin{figure*}[!hbt]
    \centering\includegraphics[width=0.82\textwidth]{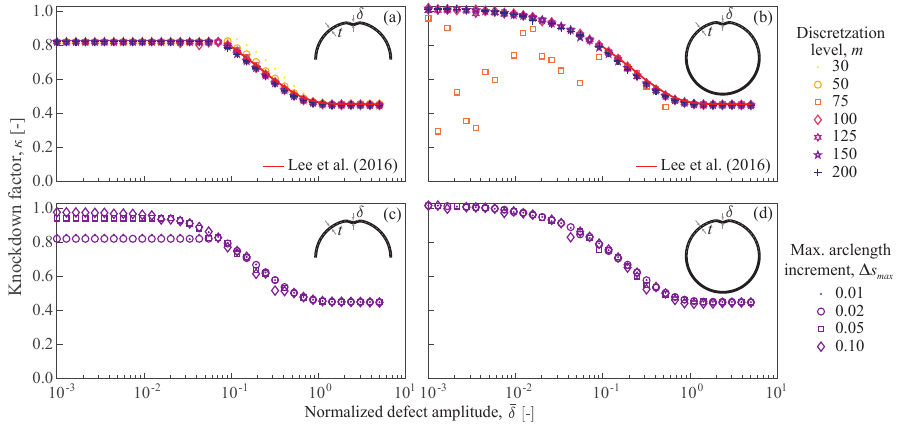}
    \caption{Sensitivity analysis of discretization and arclength-solver parameters on the buckling strength of imperfect spherical shells. (a,\,b) Imperfection sensitivity curves ($\kappa(\bar{\delta})$) for different levels of discretization $m$ (see the legend, common to both panels) and maximum arclength increment $\Delta s_{max}=0.02$; (a)  hemispherical shell clamped at the equator and (b) full spherical shell. (c,\,d) Imperfection sensitivity curves ($\kappa(\bar{\delta})$) for different maximum arclength increment $\Delta s_{max}$ (see the legend, common to both panels) and discretization level $m = 150$; (c) hemispherical shell clamped at the equator and (d) full spherical shell.
    \label{fig2_senstivity}}
\end{figure*}

To examine the impact of discretization on imperfection sensitivity, we analyzed varying discretization levels for hemispherical and full spherical shell models, as detailed in \S \ref{sec_FEM}. 
Figures~\ref{fig2_senstivity}(a,\,b) show imperfection-sensitivity curves ($\kappa$ \textit{vs.} $\bar{\delta}$) across a range of discretization levels $m$ (see legend); the results for a hemispherical shell are shown in panel (a), and for a full spherical shell in panel (b).
For hemispherical shells, convergence was achieved when $m > 75$, with consistent-imperfection sensitivity trends observed overall, though minor deviations from the trend appeared in the near-perfect limit at $m=100$. In contrast, full spheres required finer discretization, as models with $m = 75$ failed to converge for normalized defect amplitudes $\bar{\delta} < 0.1$, and small deviations were observed even at $m=125$ in the near-perfect limit. Consequently, we determined that convergence in the near-perfect limit for both hemispheres and full spheres is achieved satisfactorily with $m \geq 150$, and we will use this value for the remainder of the analysis.

With sufficient discretization,  two distinct behaviors were observed in the near-perfect limit: for full spheres, the knockdown factor approaches 1 as $\bar{\delta} \rightarrow 0$, while for hemispheres, it plateaus around 0.82. At this stage of the study, it remains to be clarified whether this difference arises from numerical artifacts or represents a fundamental physical behavior; this discrepancy will be discussed in detail in \S \ref{sec_boundary_modes}.

In the elastic buckling regime, imperfection sensitivity is expected to be independent of the material model, which is confirmed by comparing the present results for a linear elastic material model (with $E = 210\,$GPa and $\nu = 0.3$) with earlier results from Ref.~\cite{lee_fabrication_2016}. The latter are plotted as the solid red line in Fig\,\ref{fig2_senstivity}(a,\,b), which used a hyperelastic NeoHookean material (with $E = 1.26\,$GPa and $\nu = 0.5$) with the same radius-to-thickness ratio, $R/t = 100$, as that of the present study. The excellent agreement between these two sets of results (present and past) provides a quantitative verification of the model and additional confidence in its accuracy.

Having established an appropriate discretization level ($m > 150$), we proceed to investigate the sensitivity of one of the Static/Riks solver parameters, the maximum arclength increment $\Delta s_{max}$. For these simulations, the initial and minimum arclength increments were fixed at $\Delta s_{ini} = 0.01$ and $\Delta s_{min} = 10^{-10}$, respectively, while $\Delta s_{max}$ was varied between 0.01 and 0.1. Figures \ref{fig2_senstivity}(c,\,d) show the imperfection sensitivity curves ($\kappa(\bar{\delta})$) for hemispherical shells, in panel (c), and full spherical shells, in panel (d), across these maximum admissible increment values $\Delta s_{max}$ (see legend), with a fixed discretization level of $m = 150$. 
While the maximum admissible arclength increment did not impact the convergence behavior of sufficiently discretized full spheres, it had a significant effect on hemispheres, particularly in the near-perfect limit. For $\Delta s_{max} \geq 0.05$, the knockdown factor for hemispheres approached 1, similar to that of full spheres, whereas for $\Delta s_{max} \leq 0.02$, the knockdown factor reached a plateau in the near-perfect limit. These distinct trends in the near-perfect limit call for further investigation into the buckling modes in this regime.  

As we will evidence in the next section, the observed difference of two distinct knockdown factor trends in the near-perfect limit is attributed to the distinct buckling modes exhibited by spherical shells. The choice of the maximum arclength increment dictates which post-buckling branch is followed in the near-perfect shells where the boundary modes and the localized modes are close to each other. 

\section{Boundary modes versus the localized buckling mode}
\label{sec_boundary_modes}

Consistently with previous studies~\cite{lee_geometric_2016, hutchinson_buckling_2016, jimenez_technical_2017, abbasi_comparing_2023}, our results in Fig.~\ref{fig2_senstivity} show that the knockdown factor decreases and reaches a plateau as the defect amplitude $\bar{\delta}$ increases, with knockdown factors becoming independent of defect size for sufficiently large imperfections. However, in the opposite limit of near-perfect hemispherical shells ($\bar{\delta}\rightarrow 0$), the knockdown factors appear to stagnate around a finite value of less than 1, a trend highly dependent on the maximum admissible arclength increment (as shown in Fig.~\ref{fig2_senstivity}c). This sensitivity was not observed in full spherical shells (cf. Fig.~\ref{fig2_senstivity}d), suggesting that boundary conditions play a role in the buckling response of hemispheres.

\begin{figure}[!hbt]
   \centering\includegraphics[width=0.95\linewidth]{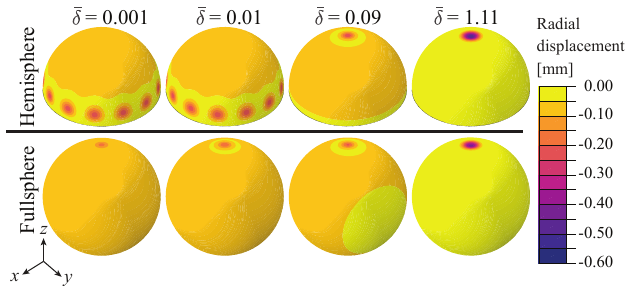}
\caption{Buckling modes of the near-perfect hemispherical shells (upper row) and full spherical shells (lower row) for $\bar{\delta} = \{0.001, 0.01, 0.09, 1.1\}$. A boundary mode is present in the clamped hemispherical shells for $\bar{\delta} = \{0.001, 0.01\}$. 
\label{fig_bucklingmode}} 
\end{figure}

Figure~\ref{fig_bucklingmode} shows visualizations snapshots of the buckling modes of hemispherical and full spherical shells for representative examples with normalized defect amplitudes of $\bar{\delta}=\{ 0.001,\, 0.01.\, 0.09,\, 1.11\}$. For all these simulations, the maximum arclength increment was set to 0.02 and the discretization level to $m = 150$. The snapshots reveal a clear distinction between the buckling behaviors of the two geometries. The hemisphere develops a non-axisymmetric, periodic boundary buckling mode along the clamped equator for $\bar{\delta}=\{ 0.001,\, 0.01.\}$ and an axisymmetric localized buckle at the pole when $\bar{\delta}=\{0.09,\, 1.11\}$. By contrast, the full sphere always exhibits only 
the axisymmetric localized buckle at the pole for all values of $\bar{\delta}$. This contrasting behavior underscores the influence of boundary conditions, highlighting that results in the near-perfect limit cannot be universally applied across shell geometries. The boundary conditions in hemispheres drive a different imperfection sensitivity compared to full spheres, making generalization across shell types unfeasible. 
Note that the early work by Budiansky and Huang on clamped shallow spherical shells has also shown that the emergence of a non-axisymmetric bucking mode near the clamped edge leads to a considerable reduction in the knockdown factor \cite{budiansky_buckling_1959,huang_unsymmetrical_1964}. 

\section{Effect of radius-to-thickness ratio}\label{sec_effect_of_eta}

\begin{figure}[!hbt]
\centering{\includegraphics[width=0.8\linewidth]{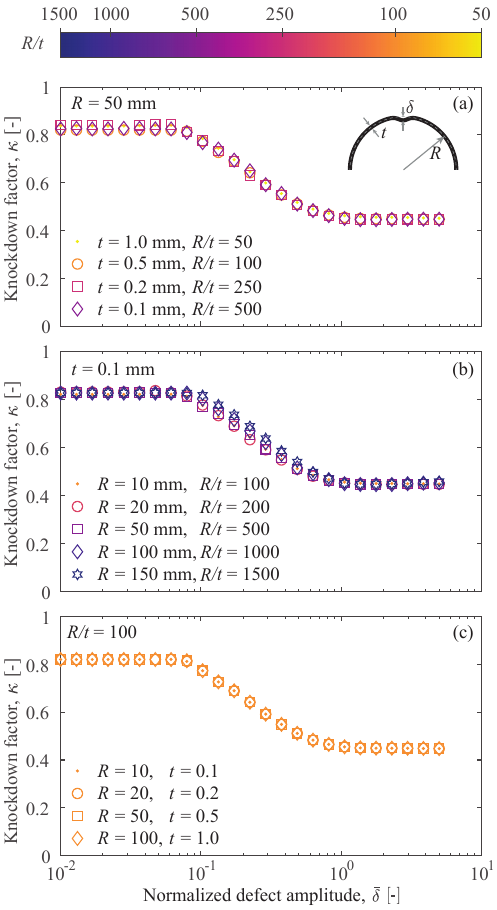}}
\caption{Knockdown factor vs. normalized defect amplitude for various shell geometry ranging for different radius-to-thickness ratios in the range $50 \leq R/t\leq 1500$: (a) constant $R=50\,$mm and varying $t$, (b) constant $t=0.1\,$mm and varying $R$, and (c) constant $R/t=100$ by varying both $R$ and $t$. See the legends for the values of the varied parameters.
\label{fig_Rteta}}
\end{figure}

It is also important to establish the generality of our results on imperfection sensitivity of the buckling of hemispheres for different radius-to-thickness ratios, $R/t$. For this purpose, we conducted simulations across in the range of $50 \leq R/t\leq 1500$, achieved by adjusting the radius and thickness combinations. These simulations were performed in three different scenarios: (1) keeping the mid-surface radius constant while varying the thickness (Fig.\ref{fig_Rteta}a), (2) holding thickness constant while varying the radius (Fig.\ref{fig_Rteta}b), and (3) maintaining a constant radius-to-thickness ratio, $R/t$ (Fig.~\ref{fig_Rteta}c). All three scenarios yielded consistent imperfection sensitivity trends, reinforcing that $R/t$ serves as a reliable parameter governing the buckling behavior of spherical shells across a range of geometric configurations. As shown in Fig.~\ref{fig_Rteta}, the knockdown factor trends remained consistent across the full explored range of $R/t$, with an upper bound plateau emerging for normalized amplitudes below $\bar{\delta}\leq 0.07$. The value of $\kappa$ of this plateau coincides with that below the transition in the buckling mode discussed in \S\ref{sec_boundary_modes}, associated with the shift from a localized dimple at the pole to a non-axisymmetric boundary mode along the equator. 

\section{From shallow to deep and full spherical shells}
\label{sec_shallow}

\begin{figure}[!t]
\centering{\includegraphics[width=\linewidth]{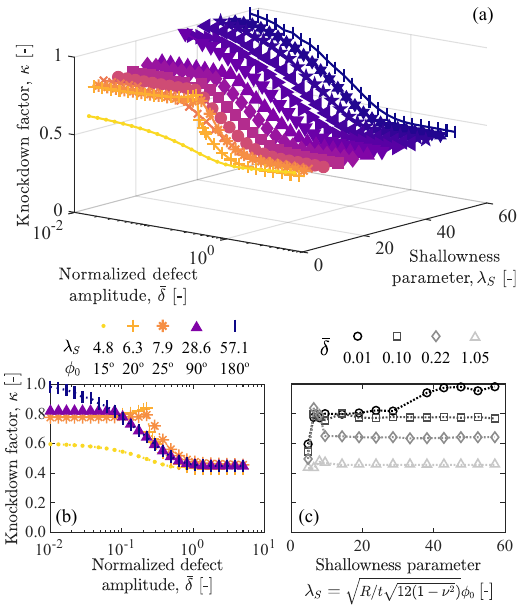}}
\caption{(a) Knockdown factor vs. normalized defect amplitude for the shell geometries with various cap half-angle $\phi_0$ ranging from 15$^o$ to 180$^o$. (b) Knockdown factor vs. normalized defect amplitude $\overline{\delta}$ for selected shallowness parameters. (c) Knockdown factor vs. shallowness parameter $\lambda_S$ for selected normalized defect amplitudes.
\label{fig_shallow}}
\end{figure}

\begin{figure*}[!hbt]
\centering\includegraphics[width=0.95\linewidth]{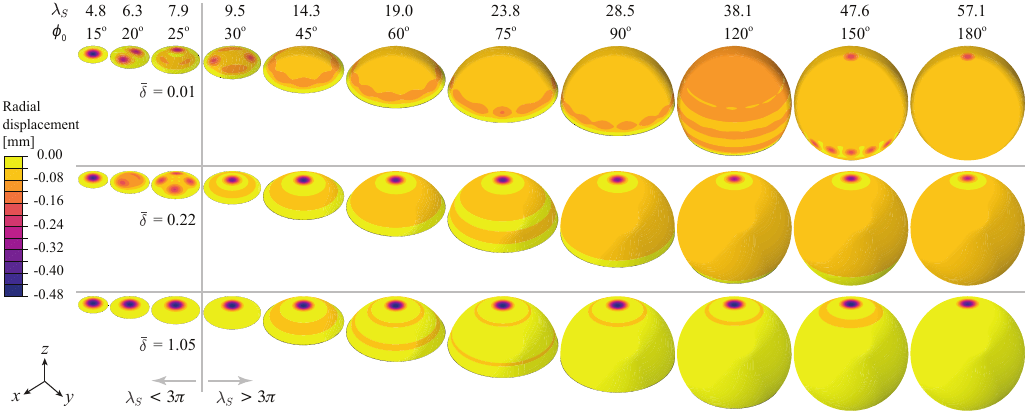}
\caption{Buckling modes of various partial spherical shells for $\bar{\delta}$ = 0.01, 0.22, 1.05. The boundary mode is present in the clamped partial spherical shells for low defect amplitudes.
\label{fig_shallow_modes}} 
\end{figure*}

Building on the observed mode-switching behavior in hemispherical shells, we extended our exploration to partial spherical shells with varying cap half-angles, $\phi_0$. Figure \ref{fig_shallow}(a) presents imperfection sensitivity curves, $\kappa(\bar{\delta})$ across a range $\phi_0$, but plotted as a function of the shallowness parameters, $\lambda_S$ via Eq.~(\ref{Eq_lambdaS}). The corresponding snapshots for selected values of $\phi_0$ (\textit{i.e.}, $\lambda_S$) and $\bar{\delta}$ are shown in Fig.~\ref{fig_shallow_modes}.
For shells with $\phi_0 < 90^\circ$ ($\lambda_S<28.6$), the nonaxisymmetric, periodic, boundary-dominated buckling modes were observed for sufficiently small dimple imperfections. When these boundary modes are excited, the $\kappa(\bar{\delta})$ curves exhibit the upper plateau that was already discussed in \S\ref{sec_boundary_modes} for hemispherical shells. Notably, for shells with cap half-angles lower than $\phi_0 < 30^\circ$ ($\lambda_S<9.5$), the transition between buckling modes, from the periodic mode near the boundary to the localized mode at the pole, occurred at larger defect amplitudes compared to other partial shells. By contrast, for $\phi_0 \gtrsim 90^\circ$ ($\lambda_S \gtrsim 28.6.3$, the `$\gtrsim$' symbol being used given the resolution of the steps in the values of $\phi_0$ of the explored data), the knockdown factor of all shells tends to unity ($\kappa\rightarrow 1$) in the limit of near-perfect geometry ($\bar{\delta}\rightarrow 0$).  

In Fig.~\ref{fig_shallow}(b), we replot $\kappa(\bar{\delta})$ curves for partial shells with a selection of cap half-angles values: $\phi_0{=}\{15^\circ,20^\circ,25^\circ,90^\circ,180^\circ\}$. This representation of the data further clarifies the two general groups of curves, one for deep shells where $\kappa\rightarrow\approx1$ as $\bar{\delta}\rightarrow 0$, and the other for shallow shells with an upper plateau of $\kappa\approx0.8$ as $\bar{\delta}\rightarrow 0$. The very shallow shell with $\phi_0 \approx 15^\circ$ (\textit{i.e.}, $\lambda_S=4.8$) is an exception in the latter group presumably because the characteristic length scale of the boundary mode is comparable to and competes strongly with the system size, as suggested by the corresponding snapshot in Fig.~\ref{fig_shallow_modes}. 
In Fig.\,\ref{fig_shallow}(c), the knockdown factor is plotted against the shallowness parameter $\lambda_S$ for selected constant normalized defect amplitudes $\bar{\delta}$. It should be pointed out that the results for the nearly-perfect shallow shells ($\bar{\delta}=0.01$ and $\phi_0 < 30^\circ$) shown in Fig.\,\ref{fig_shallow}(c) are in good agreement with the analysis by Huang \cite{huang_unsymmetrical_1964}, which highlighted the significant reduction in buckling pressure produced by asymmetric deformation. 

As mentioned earlier, for shallow shells with lower $\lambda_S$ values, symmetric buckling dominates, and as $\lambda_S$ increases, the buckling transitions to asymmetric modes. Earlier studies demonstrated that the transition from symmetric to asymmetric buckling occurs when $\lambda_S > 5.5$. \cite{parmerter_influence_1962,parmerter_buckling_1964, huang_unsymmetrical_1964, thurston_asymmetrical_1964}. Figure~\ref{fig_shallow_modes} illustrates this transition, showing that for near-perfect shells with $\overline{\delta} = 0.01$: the mode does indeed shift from symmetric to asymmetric around $\lambda_S \approx 5.5$.
For intermediate defect amplitudes ($0.1 \leq \overline{\delta} \leq 0.6$), the knockdown factor exhibits non-monotonic behavior, reflecting the complex interplay between the defect size and the shallowness parameter. By contrast, for larger defect amplitudes, the knockdown factor becomes largely insensitive to variations in $\lambda_S$, indicating that geometric imperfections dominate the buckling response in this regime.

Figure~\ref{fig_shallow_modes} shows typical snapshots of the post-buckling modes, from shallow to deep shells, for selected defect amplitude values. The main difference in the behavior is attributed to the limit $\lambda_S\approx 3\pi=9.4...$ (\textit{i.e.}, $\phi_0\approx 29.7^\circ$) suggested by Hutchinson~\cite{hutchinson_imperfection_1967} that was already discussed in \S\ref{sec_probdef}; beyond this limit, the spherical cap is more than 3 times larger than the theoretical buckling wavelength $l_c$. From the snapshots in Fig.~\ref{fig_shallow_modes}, it is evident that, in the neighborhood of this limit, there is a transition between the boundary-dominated non-axisymmetric and the localized axisymmetric buckling modes, for intermediate values of $\bar{\delta}=0.22$. For $\bar{\delta}=0.01$, this transition happens when $\phi_0 > 90^\circ$.
Note that for even smaller cap half-angles ($\phi_0 < 10^\circ$), the shells exhibited plate-like bending, with no snap-through buckling, corresponding to a shallowness parameter $\lambda_S < \pi$; a finding aligning with previous results in Ref.~\cite{ki_combined_2024}.
To contextualize this shallowness limit, the buckling wavelength's half-angle is approximately 10$^\circ$ (equivalent to an arclength of $0.5l_c/R$).
When the arclength of the partial shell ($2R\phi_0$) falls below the full buckling wavelength, the shell effectively behaves as a plate under bending, and snap-through buckling is no longer initiated. 
For intermediate shallowness parameters ($\pi\leq\lambda_S\leq 3\pi$), larger defect amplitudes produced higher knockdown factors, indicating that buckling behavior in shallow spherical caps is heavily influenced by both geometric shallowness and defect size.

\section{Discussion and Conclusion}
\label{sec_conclusion}

We have systematically investigated the buckling of imperfect spherical shells through FEM simulations, examining the effects of discretization and solver settings, as well as geometric parameters of both the defect and the overall shell shape. We focused on Gaussian-dimple defects, particularly in the near-perfect limit of vanishing defect amplitudes.

First, we conducted a sensitivity analysis on discretization and solver parameters to ensure the robustness and accuracy of FEM-computed results. Naturally, we found that convergence in knockdown factor computations depends strongly on the discretization level, particularly in the near-perfect limit of interest. If the mesh is too coarse, its discrete approximation can act as an unintended imperfection that dominates the actual geometric defect, yielding spurious results. The solver parameters are also critical. We varied the arclength increment in the Static/Riks method; while there was minimal impact on full spherical shells, we found a significant influence on hemispherical shell buckling. These results demonstrate that FEM simulations of idealized spherical shells require careful handling of imperfections and solver parameters, especially in the near-perfect limit due to the proximity of multiple buckling branches.
Consequently, our recommendations for imperfection-sensitive buckling analysis of spherical shells are:
\begin{enumerate}
\item Gaussian dimple imperfections should be introduced by gradually reducing their amplitude to approach the near-perfect limit; this minimizes the solver and discretization sensitivity while providing a realistic assessment of imperfection effects.
\item The theoretical buckling load, $p_c$, in Eq.~(\ref{Eq_Gaussian}) should be used as the nominal load for the arclength solver, given that the increment parameters depend directly on the applied load.
\item To quantify knockdown factors, the measured/simulated critical buckling loads should be normalized by the theoretical buckling load, $p_c$, of a full sphere, as the simulated response of the near-perfect geometry is highly sensitive to use as a normalization factor.
\end{enumerate}
Future work should tackle nearly perfect non-spherical shells for which no closed-form theoretical solution exists for the buckling strength of the perfect geometry, thereby comprising the recommendations (2) and (3) above. 


Once the appropriate discretization and solver parameters were determined, we then checked the generality of our results in a range of radius-to-thickness ratios, $50\leq R/t\leq1500$. Across this wide range, the imperfection sensitivity trends and knockdown factors remained consistent. Additional simulations with fixed radius, thickness, or $R/t$ values confirmed that this radius-to-thickness ratio governs the imperfection sensitivity of thin spherical shells. 

From a mechanics viewpoint, our key finding is the identification of distinct buckling modes between hemispherical and full-spherical shells. In the near-perfect limit ($\bar{\delta}\rightarrow0$), hemispherical shells exhibit boundary-dominated, non-axisymmetric, periodic buckling modes along the clamped equator, while a localized axisymmetric buckling response concentrated around the pole always dominates full spherical shells. Moreover, in the $\bar{\delta}\rightarrow0$ limit, the knockdown factor of hemispherical shells reaches an upper-bound plateau of $\kappa\approx0.8$, whereas full spherical shells exhibit the expected $\kappa\rightarrow 1$ behavior when $\bar{\delta}\rightarrow0$. This upper-bound plateau for the hemispheres is found for defect amplitudes below $\bar{\delta} < 0.07$. The reason for this intriguing behavior of the hemispherical shells is that, when they are near-prefect, the boundary modes become the \textit{de facto} imperfection dominating over the Gaussian dimple at the pole. By contrast, for the full spherical shells, in the absence of boundary modes, the Gaussian dimple rules throughout. This difference in buckling modes between the two cases suggests that boundary conditions significantly influence imperfection sensitivity and knockdown factors. 

We also explored systematically shells with different geometries by varying the cap half-angle $\phi_0$, which relates to the shallowness parameter $\lambda_S$ through Eq.~(\ref{Eq_lambdaS}). The threshold between the two qualitatively distinct responses overviewed in the previous paragraph occurs at $\phi_0 < 90^\circ$ (\textit{i.e.}, $\lambda_S<28.6$), with boundary-dominated buckling ceasing to be observed for deeper shells beyond this value. Furthermore, when $\phi_0 < 10^\circ$ (\textit{i.e.}, $\lambda_S<\pi$), the buckling response of these very shallow shells is more akin to plate bending, with no snap-through behavior. 
Additionally, as demonstrated by earlier studies \cite{parmerter_influence_1962, huang_unsymmetrical_1964, thurston_asymmetrical_1964}, the transition from symmetric to asymmetric buckling occurs at $\lambda_S \approx 5.5$ for the near-perfect shells. However, this buckling mode transition becomes less apparent in the presence of significant geometric defects. 
Overall, this analysis suggests four distinct regimes based on $\lambda_S$ and $\overline{\delta}$: (1) $\lambda_S < \pi$, where no snap-through buckling occurs; (2) $\pi \leq \lambda_S \leq 5.5$, where the buckling mode is symmetric and then transitions to an asymmetrical mode in the limit of $\bar{\delta}\rightarrow0$; (3) $5.5 \leq \lambda_S \leq 3\pi$, where boundary and localized modes interact with the former dominating in lower defect amplitudes  $\bar{\delta}$; and (4) $\lambda_S \geq 
3\pi$, where a localized, axisymmetric buckling mode near the pole of the shell dominates the buckling behavior.

\section*{Acknowledgment} 

The authors are grateful to John Hutchinson for the helpful discussions and suggestions.


\bibliographystyle{asmejour}   
\bibliography{refs} 

\end{document}